# The Orbital Period and Variability of the Dwarf Nova ES Draconis

F. A. Ringwald and Kenia Velasco

Department of Physics
California State University, Fresno
2345 E. San Ramon Ave., M/S MH37
Fresno, CA 93740-8031, U. S. A.

E-mail: `ringwald@csufresno.edu`

ABSTRACT

A radial velocity study of the cataclysmic variable ES Dra (PG 1524+622) is presented. ES Dra is found to have an orbital period of 0.17660 ± 0.00006 d (4.2384 ± 0.0014 h). The mass-losing secondary star of ES Dra is detectable in the spectrum, and it has a spectral type of M2 ± 1. From this, we estimate the absolute magnitude of ES Dra during our spectroscopic observations to have been $M_R$ = 6.5 ± 0.5, and its distance to be 720 ± 150 pc. The long-term light curve of ES Dra compiled by the American Association of Variable Star Observers (AAVSO) shows that ES Dra is a Z Cam star, which between 1995 and 2009 spent most of its time in standstill.

## 1. INTRODUCTION

Cataclysmic variable binary stars (CVs) are close binary star systems, in which typically a late-type, approximately main-sequence star, called the secondary star, orbits a white dwarf. In all CVs, the secondary star fills its Roche lobe, or gravitational equipotential, and spills gas onto the white dwarf. Because the two stars are orbiting each other, the gas stream does not strike the white dwarf directly: it forms a ring around the white dwarf, which viscosity in the gas broadens into an accretion disk, through which the gas accretes onto the white dwarf. Reviews on CVs include those by Robinson (1976), Warner (1995), and Hellier (2001).

CV orbital periods ($P_{orb}$) range from about 78 minutes to 12 hours, or longer. Many properties of CVs depend on their orbital periods, including their secular evolution, luminosity, and outbursts (Shafter, Wheeler, & Cannizzo 1986). The catalogs of Patterson (1984) and Ritter & Kolb (2003) include only CVs of known or suspected orbital period, emphasizing this parameter's importance.

Radial velocity studies reveal orbital periods in CVs more reliably than do photometric modulations, save for eclipses, but are not without problems (Araujo-Betancor et al. 2005). The strongest features in CV spectra are their emission lines, which originate in the CVs' accretion disks. Orbital amplitudes, or $K$–velocities, when measured from the emission lines, do not reliably trace the motion of the white dwarf, since they originate in the disk. Neither do $\gamma_{em}$, the emission line mean velocity, nor $T_0$, the epoch of spectroscopic phase zero. Eclipsing systems sometimes show $T_0$ lagging the eclipse by over 70º in phase (e.g., Thorstensen et al. 1991).

This paper reports a spectroscopic period measurement of the cataclysmic variable ES Draconis. ES Dra was discovered by the Palomar-Green survey (Green, Schmidt, & Liebert 1986) and listed as PG 1524+622. In Section 2, observational procedure, data reduction, and analysis are described. The average optical spectrum of ES Dra is discussed in Section 3. We estimate the absolute magnitude and distance of ES Dra in Section 4. The long-term light curve and outburst behavior of ES Dra are discussed in Section 5, with conclusions in Section 6.

2. OBSERVATIONS

Spectra of ES Dra were taken with the Mark IIIa spectrograph on the 1.3-m McGraw-Hill telescope at MDM Observatory. The long-slit spectra permitted accurate sky subtraction, although absolute spectrophotometry may have been accurate only to within 30% because of losses through the 2.2-arcsecond slit that was used for all exposures. Instrument rotation was deemed unnecessary for such a red setup. The detector was a TI-4849 CCD in the BRICC camera (Luppino 1989).

A 300 line/mm grism blazed at $\lambda$ 6400 Å was used with a Hoya Y-50 order-sorting filter in the spectrograph, important since CVs have very blue continua. The wavelength coverage spanned about $\lambda\lambda$ 6200 – 9000 Å at 11-Å resolution (at 5 Å per channel dispersion). Exposure times for all spectra of ES Dra were 900 seconds.

All observations were made in photometric conditions. Flux standards from Oke (1974) were taken at least once per night, to calibrate the instrumental response. Spectra of F8 stars, hot enough to be nearly featureless in the red and common enough to be always closer than 5 degrees from ES Dra, were also taken, for mapping and removing atmospheric absorption bands (see Friend et al. 1988).

All spectra were reduced and analyzed as described by Thorstensen & Freed (1985) and Ringwald, Thorstensen, & Hamwey (1994). Radial velocities were measured by convolving the Hα emission line with two Gaussians that were 2.8 channels (640 km s$^{-1}$) wide, separated by 1300 km s$^{-1}$ (see Shafter 1983). Table 1 lists the measured velocities.

A Lomb-Scargle periodogram (Lomb 1976; Scargle 1982; Press et al. 1992) of the radial velocities is shown in Figure 1. The correct alias is obvious, at a period of 0.17660 ± 0.00006 days. Both statistics from the Monte Carlo analysis of Thorstensen & Freed (1985), the discriminatory power and the correctness likelihood, are over 99.9%.

Figure 2 shows a least-squares fit to a sinusoid with the most likely period. Spectrograph flexure caused variations of ± 10 km s$^{-1}$, shown by velocities measured from the night-sky line at λ 5577 Å and plotted in Figure 2. Table 2 lists the derived orbital parameters.

3. SPECTRUM

An average of the individual spectra, comprising 8.25 hours' observing time, is shown in Figure 3. This spectrum of ES Dra is typical for a dwarf nova not at outburst maximum, with Balmer lines in strong, broad emission, a flat continuum, and TiO bands at λλ 6800, 7150, and 7600 Å.

We scaled, fitted, and subtracted spectra of a variety of red dwarfs from this spectrum, as described (for BZ UMa) by Ringwald, Thorstensen, & Hamwey (1994). We estimate the spectral type of the mass-losing secondary star in ES Dra to be an M2 ± 1, since an M2 dwarf gave the smoothest continuum after this scaling and subtraction. An M2 type is consistent with the orbital period derived in this paper (see Figure 7 of Knigge 2006).

The spectrum of ES Dra (Figure 3; see also Figure 10 of Downes et al. 1995) shows only relatively low-excitation emission lines. There is no He II λ 4686 Å emission, characteristic of CVs with magnetic white dwarfs, and of novae with recent eruptions (see Williams 1983; see also Figure 3.4 of Hellier 2001). Hα is in strong emission, with an equivalent width of 40 ± 5 Å and a full-width at zero intensity of 5300 ± 200 km s$^{-1}$. There is no sign of the absorption wings that dwarf novae in outburst maximum or nova-like variables often show (see Williams 1983; see also Figure 3.6 of Hellier 2001). Some of the Paschen lines and He I λ 6678 Å are in weak emission. O I λ 7773 Å is barely noticeable above the noise level, if present at all (see Friend et al. 1988). The spectrum shows both emission lines and a continuum that allows the mass-losing secondary star to show through, so it is consistent with ES Dra being a dwarf nova that was not in outburst maximum when these observations were made (Williams 1983).

Misselt & Shafter (1995) collected time-resolved photometry during two nights, spanning one hour on one night and just over one full orbit on the other night. No eclipses were obvious, and the time resolution of 40-50 seconds did not reveal any other obvious photometric variations, such as coherent oscillations from a magnetic white dwarf (see Warner 1995). This and the absence of high-excitation emission lines in the spectrum show that ES Dra is probably not a magnetic CV.

## 4. ABSOLUTE MAGNITUDE AND DISTANCE

From the average spectrum (Figure 3), we estimate a through-the-slit magnitude for ES Dra at the time the spectra were taken to have been $R = 15.8 \pm 0.2$. We also estimate the fraction of the light of the mass-losing secondary star of ES Dra at $\lambda$ 7000 Å to have been $18 \pm 3$ %. If the secondary star is a main-sequence M2 ± 1, with $M_V = +9.9$ and $V - R = 1.50$ for an M2 dwarf (p. 388 of Cox 2000), and also assuming that $E(B - V) = 0$ and $V - R \approx 0$ for ES Dra, this implies that ES Dra had $M_R = +6.5 \pm 0.5$ at the time of our spectroscopic observations. It also implies that ES Dra has a distance $d = 720 \pm 150$ pc.

This does not agree with the estimate of $d = 1358$ pc of Ak et al. (2008), which was based on their period-luminosity-color relation for infrared magnitudes. It was also based on their estimate of $E(B-V) = 0.018$ for ES Dra, which was calculated from the maps of interstellar dust of Schlegel, Finkbeiner, & Davis (1998) and the model of interstellar absorption of Bahcall & Soneira (1980) along the line of sight to ES Dra, which is at $b = 47º$. We calculate that the recalibration of Ramseyer (1994) of the method of Bailey (1981) gives a lower limit for the distance to ES Dra of $530 \pm 70$ pc, assuming our orbital period and the 2MASS magnitudes for ES Dra of $J = 15.458 \pm 0.064$, $H = 14.880 \pm 0.077$, and $K_s = 14.889 \pm 0.127$ (Cutri et al. 2003), with the transformation of Carpenter (2001) to the C. I. T. system (Elias et al. 1982).

Our distance estimate is consistent with the best estimate of Godon et al. (2009) of $d = 770$ pc, which is based on the estimate of 702 pc from the outburst maximum versus orbital period relation of Warner (1987), the estimate of 840 pc from the similar but recalibrated relation of Harrison et al. (2004), and the estimate of 540-1016 pc from infrared magnitudes and models of secondary stars of Knigge (2006, 2007). Godon et al. (2009) also measured $E(B-V) = 0.02$, from continuum fits to their ultraviolet spectrum of ES Dra. Misselt & Shafter (1995) showed that ES Dra is not eclipsing, but this only constrains the orbital inclination to $i \leq 70º$. The inclination correction to the absolute magnitude of Warner (1987) may therefore be negligible. We therefore adopt our own distance estimate of $720 \pm 150$ pc in the discussion below.

## 5. LONG-TERM LIGHT CURVE

Andronov (1991) examined 57 archival plates and found variability between $m_{pg} = 13.9$ and 16.3, with one variation of 1.3 magnitudes in one day. A long-term light curve of observations by the American Association of Variable Star Observers (AAVSO) (Henden 2009), from 1995 October to 2009 December, is shown in Figure 4. This light curve includes both $V$ magnitudes measured by CCD imaging and visual magnitudes ($m_{vis}$) by eyepiece observers: we expect the difference between $V$ and $m_{vis}$ to be small, relative to the ± 0.1-magnitude errors typical for eyepiece visual magnitude estimates. Figure 4 does not include apparent magnitudes measured through other filters, nor does it include magnitude limits of non-detections.

Dwarf nova outbursts are obvious in Figure 4, but they have relatively small amplitudes of 1-2 magnitudes brighter than the apparent magnitude at which ES Dra spends most of its time, $V \approx 15.1$-15.5. Notice also the fadings to $V = 17$ or fainter, which go considerably fainter than the archival plates examined by Andronov (1991).

One interpretation of this long-term light curve may be that ES Dra is a dwarf nova of the U Gem type, which is a dwarf nova that has only normal outbursts. The amplitude of the outbursts is small for a dwarf nova, though, averaging between 1 and 2 magnitudes: more typical is 2-5 magnitudes (Robinson 1976). Honeycutt, Robertson, & Turner (1998) and Honeycutt (2001) pointed out what they called "stunted" outbursts, which are sometimes present in luminous nova-like variables. The outbursts of ES Dra appear to be similar in amplitude. They do not seem to be always present. During the most active period, between JD 2,452,000 to 2,452,600 (2001 March to 2002 November, respectively), the outbursts recurred at a rate of about 7-8 outbursts per year, and lasted between 5 and 20 days, although these statistics are approximate, since coverage was sparse.

However, our spectrum of ES Dra (Figure 3) more closely resembles the relatively optically thin spectrum of a dwarf nova between outbursts. It does not show the weak emission lines and line wings in absorption that dwarf novae in outburst and nova-like variables usually show. Also, the decline in the amplitudes of the outbursts and progressive increase of faintest $V$ after 2001 March is similar to that shown by Z Cam stars going into one of their characteristic standstills (see Figure 5; see also Figure 5 of Szkody & Mattei 1984). We also note an outburst, with an amplitude of 3 magnitudes, occurring during a low state, near JD 2,453,800 (see Figure 6). We therefore think that ES Dra is a Z Cam star, and that it spent most of the time between 1995 and 2009 in standstill, not quiescence.

If ES Dra is a Z Cam star, and if its distance is $720 \pm 150$ pc, as shown above, then at outburst maximum, at $V \approx 13.9$, its absolute magnitude would be 4.6 (+0.5/-0.4). This is consistent with the relations of Warner (1987) and of Harrison et al. (2004), both of which relate absolute visual magnitude at outburst magnitude with orbital period, and which predict $M_V$ (max) = 4.5 and 4.3, respectively. If ES Dra is a Z Cam star, and if its distance is $720 \pm 150$ pc, then when $V \approx 15.1$-15.5 at standstill, $M_V = 5.8$-6.2. When $V \approx 17$ in quiescence, $M_V = 7.7$. All these values are consistent with absolute magnitudes of Z Cam stars, in their various outburst states (see Figure 11 of Warner 1987).

6. CONCLUSIONS

This paper presents a definitive orbital period of $4.2384 \pm 0.0014$ hours for ES Dra, measured by a radial velocity study. We find a spectral type for the mass-losing secondary star to be M2 ± 1, and a distance of 720 ± 150 pc. We also argue that ES Dra is a Z Cam star.

ACKNOWLEDGMENTS

The radial velocity observations were made at Michigan-Dartmouth-MIT Observatory, Kitt Peak, Arizona, which when these observations were made was owned and operated by a consortium of the University of Michigan, Dartmouth College and the Massachusetts Institute of Technology. This publication makes use of data products from the Two Micron All Sky Survey, which was a joint project of the University of Massachusetts and the Infrared Processing and Analysis Center/California Institute of Technology, funded by the National Aeronautics and Space Administration and the National Science Foundation. This research has made use of the SIMBAD database, operated at CDS, Strasbourg, France. Special thanks to Professor John Thorstensen for the use of his bubble machine, to do the Monte Carlo analysis. We acknowledge with thanks the variable star observations from the AAVSO International Database contributed by observers worldwide and used in this research.

Table 1. Hα Emission Radial Velocities for ES Dra, 1991 May 24-26[a]

| HJD[b] | V (km s$^{-1}$) | HJD[b] | V (km s$^{-1}$) | HJD[b] | V (km s$^{-1}$) |
|---|---|---|---|---|---|
| 8401.839 | −59 | 8401.963 | −173 | 8402.806 | −30 |
| 8401.850 | −31 | 8402.693 | −205 | 8402.817 | −129 |
| 8401.862 | −76 | 8402.704 | −193 | 8402.829 | −191 |
| 8401.873 | 8 | 8402.716 | −50 | 8402.840 | −232 |
| 8401.884 | 55 | 8402.727 | −62 | 8402.853 | −219 |
| 8401.895 | −34 | 8402.738 | 23 | 8402.866 | −176 |
| 8401.907 | −46 | 8402.749 | 6 | 8402.878 | −121 |
| 8401.918 | −70 | 8402.761 | 17 | 8402.889 | −106 |
| 8401.929 | −117 | 8402.772 | −2 | 8402.901 | −60 |
| 8401.940 | −122 | 8402.784 | 43 | 8402.912 | 43 |
| 8401.952 | −173 | 8402.795 | −22 | 8402.924 | 35 |

[a] Double Gaussian algorithm, separation 1300 km s$^{-1}$.
[b] Heliocentric Julian Date of mid-integration, minus 2,440,000.

Table 2. Derived Orbital Parameters for ES Dra, from Hα velocities[a]

| $P_{orb}$ (days) | $K_{em}$ (km s$^{-1}$) | $\gamma_{em}$ (km s$^{-1}$) | $T_0$ (HJD − 2,440,000) | $\sigma$ (km s$^{-1}$) |
|---|---|---|---|---|
| 0.17660 ± 0.00006 | 134 ± 11 | −75 ± 8 | 8411.024 ± 0.002 | 48 |

[a] Velocities fitted to $V(t) = \gamma_{em} + K_{em} \sin[2\pi (t - T_0)/P_{orb}]$. All measurements use the Shafter (1983) double-Gaussian method. Gaussian widths are 640 km s$^{-1}$. All errors are estimated to 68% confidence (see Thorstensen & Freed 1985).

[b] Gaussian separation 1300 km s$^{-1}$.

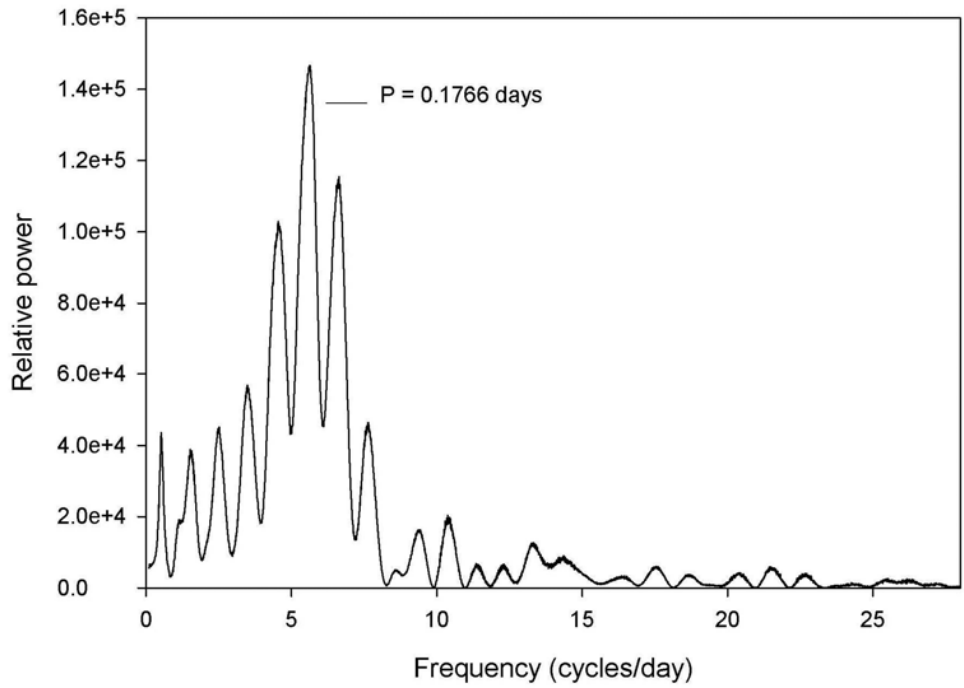

Figure 1.

A Lomb-Scargle periodogram for Hα velocities for ES Dra. The highest peak, indicating the likely orbital period to be 0.1766 days (4.24 hours), is marked.

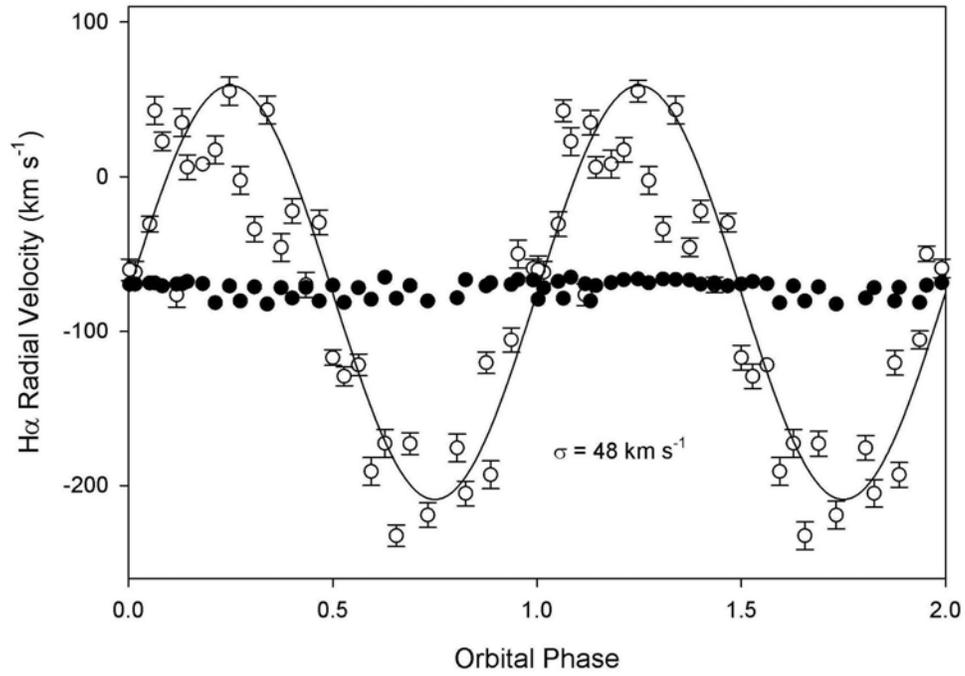

Figure 2.

A least-squares fit to a sinusoid with the most likely orbital period by the Hα velocities of ES Dra. All velocities are plotted twice for continuity, using open circles, with error bars estimated by Gaussian fits (see Ringwald et al. 1994). Also plotted with filled circles are velocities measured from the night-sky line at $\lambda$ 5577 Å, showting the amount of spectrograph flexure.

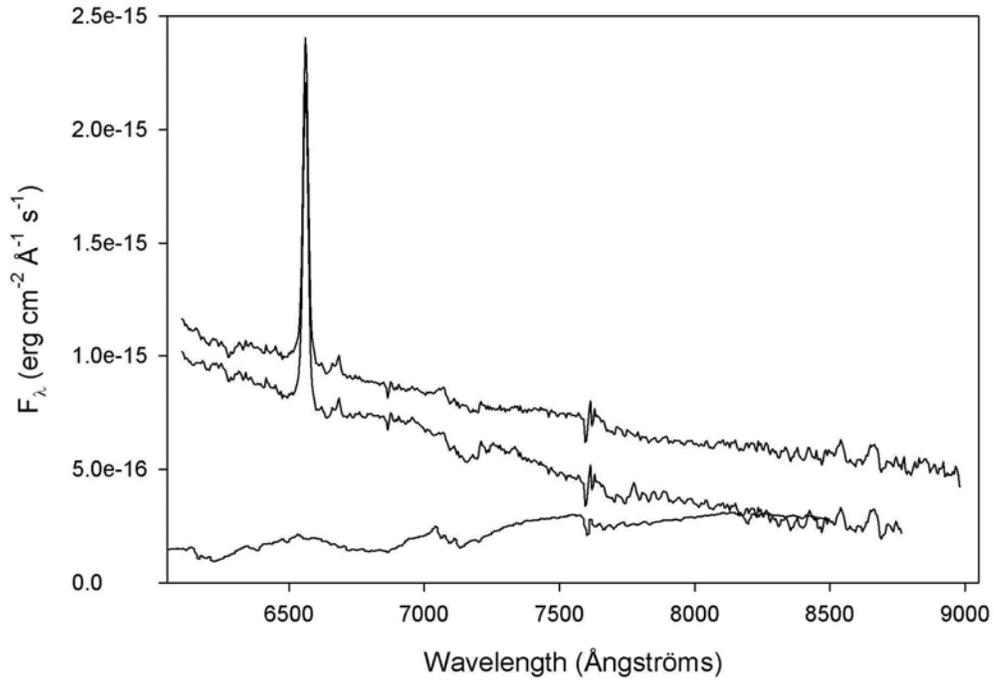

Figure 3.

(Top) Average of the spectra of ES Dra, comprising 8.25 hours' exposure time. An F8 stellar spectrum was used to map and remove the telluric absorption, with some residual absorption near $\lambda\lambda$ 6870 and 7600 Å.

(Middle) This average spectrum with a scaled M2-dwarf spectrum subtracted. This should approximate the spectrum of the disk alone.

(Bottom) A scaled M2-dwarf spectrum, taken with the same spectrograph.

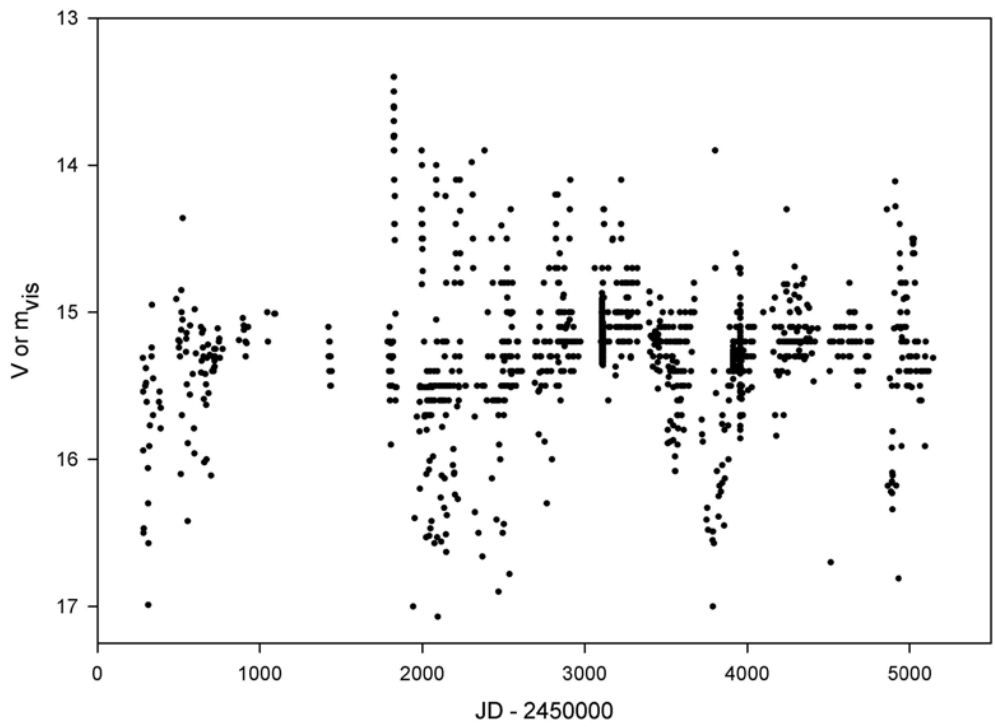

Figure 4.

A long-term light curve of ES Dra, from AAVSO observations (Henden 2009).

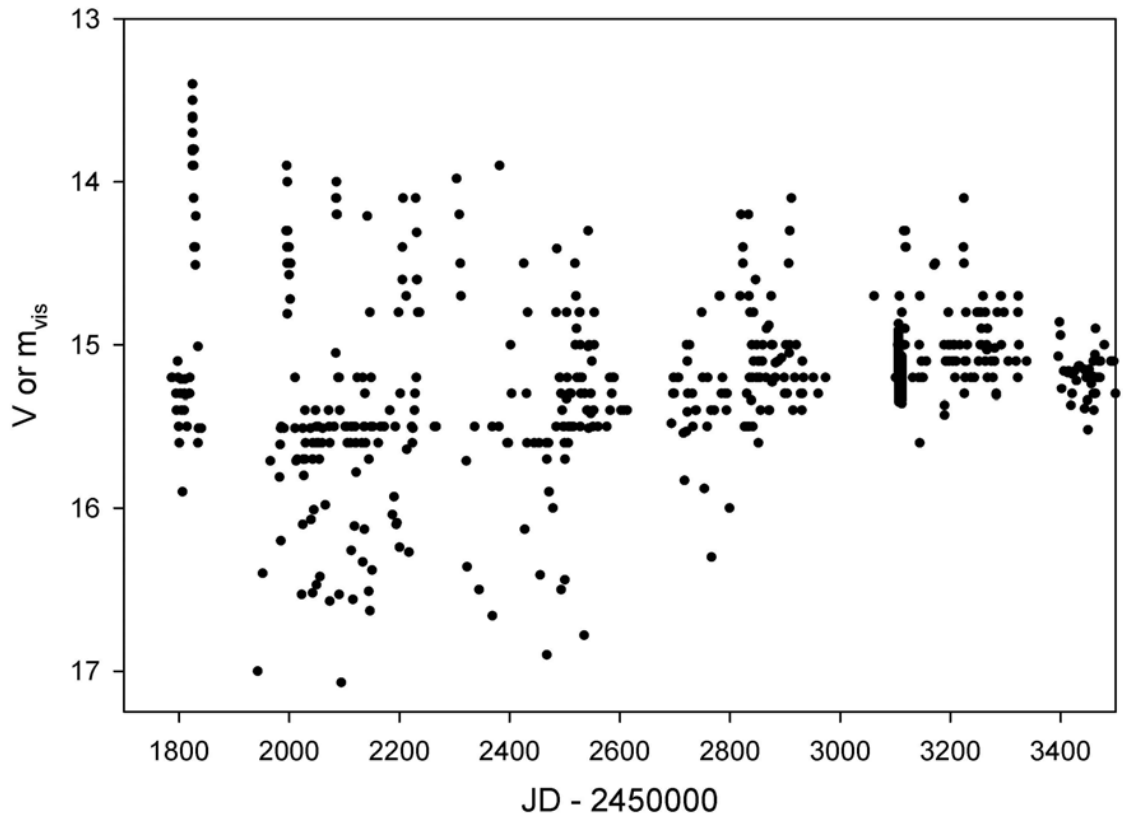

Figure 5.

A close-up of the long-term light curve of ES Dra, from AAVSO observations (Henden 2009). Notice the lessening of the amplitudes of the outbursts, and the progressive brightening of the minima between outbursts.

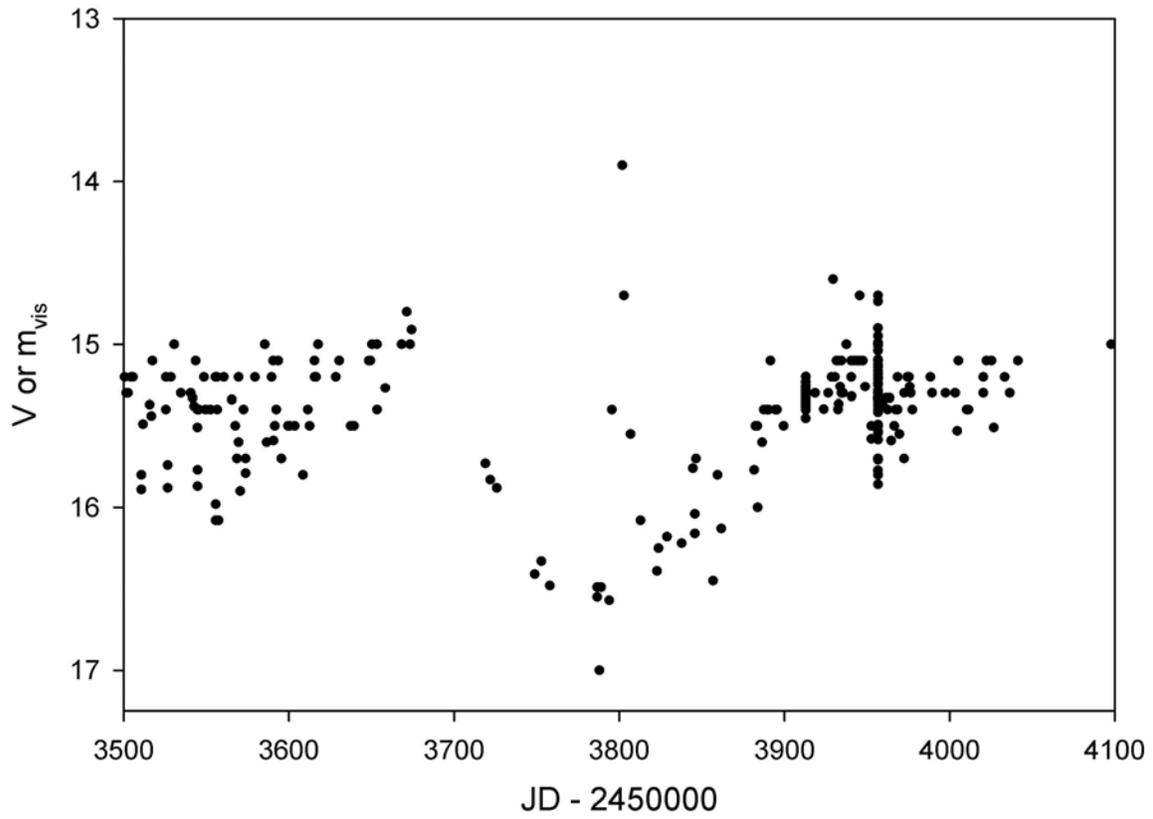

Figure 6.

A close-up of the long-term light curve of ES Dra, from AAVSO observations (Henden 2009). Notice the outburst during a minimum near JD 2,453,800.